\documentclass[%
preprint,
superscriptaddress,
 amsmath,amssymb,
 superscriptaddress
prb,
]{revtex4-1}
\usepackage[dvipdfmx]{graphicx}
\usepackage{bm}

\usepackage[top=20truemm,bottom=20truemm,left=20truemm,right=20truemm]{geometry}
\usepackage{array}
\usepackage{color}
\usepackage{setspace}
\usepackage{mathtools}

\begin{document}
\setlength\textfloatsep{11pt}

\title{Hard and soft x-ray photoemission spectroscopy study of the new Kondo system SmO thin film}

\author{Shoya Sakamoto}
\affiliation{Department of Physics, The University of Tokyo, Bunkyo-ku, Tokyo 113-0033, Japan}

\author{Kenichi Kaminaga}
\affiliation{Department of Chemistry, The University of Tokyo, Bunkyo-ku, Tokyo 113-0033, Japan}
\affiliation{WPI-Advanced Institute for Materials Research and Core Research Cluster, Tohoku University, Sendai 980-8577, Japan}

\author{Daichi Oka}
\affiliation{Department of Chemistry, Tohoku University, Sendai 980-8578, Japan}

\author{Ryu Yukawa}
\affiliation{Photon Factory, Institute of Materials Structure Science, High Energy Acceleraotor Research Organization (KEK), Tsukuba 305-0801, Japan}

\author{Masafumi Horio}
\affiliation{Department of Physics, The University of Tokyo, Bunkyo-ku, Tokyo 113-0033, Japan}

\author{Yuichi Yokoyama}
\affiliation{Institute for Solid State Physics, The University of Tokyo, Kashiwa, Chiba 277-8561, Japan}

\author{Kohei Yamamoto}
\affiliation{Institute for Solid State Physics, The University of Tokyo, Kashiwa, Chiba 277-8561, Japan}

\author{Kou Takubo}
\affiliation{Institute for Solid State Physics, The University of Tokyo, Kashiwa, Chiba 277-8561, Japan}

\author{Yosuke Nonaka}
\affiliation{Department of Physics, The University of Tokyo, Bunkyo-ku, Tokyo 113-0033, Japan}

\author{Keisuke Koshiishi}
\affiliation{Department of Physics, The University of Tokyo, Bunkyo-ku, Tokyo 113-0033, Japan}

\author{Masaki Kobayashi}
\affiliation{Photon Factory, Institute of Materials Structure Science, High Energy Acceleraotor Research Organization (KEK), Tsukuba 305-0801, Japan}

\author{Arata Tanaka}
\affiliation{Department of Quantum Matter, Graduate School of Advanced Sciences of Matter (ADSM), Hiroshima University, Higashi-Hiroshima 739-8530, Japan}

\author{Akira Yasui}
\affiliation{Japan Synchrotron Radiation Research Institute, SPring-8, Sayo, Hyogo 679-5198, Japan}

\author{Eiji Ikenaga}
\affiliation{Japan Synchrotron Radiation Research Institute, SPring-8, Sayo, Hyogo 679-5198, Japan}

\author{Hiroki Wadati}
\affiliation{Institute for Solid State Physics, The University of Tokyo, Kashiwa, Chiba 277-8561, Japan}

\author{Hiroshi Kumigashira}
\affiliation{Photon Factory, Institute of Materials Structure Science, High Energy Acceleraotor Research Organization (KEK), Tsukuba 305-0801, Japan}
\affiliation{Department of Physics, Tohoku University, Sendai 980-8578, Japan}

\author{Tomoteru Fukumura}
\affiliation{WPI-Advanced Institute for Materials Research and Core Research Cluster, Tohoku University, Sendai 980-8577, Japan}
\affiliation{Department of Chemistry, Tohoku University, Sendai 980-8578, Japan}
\affiliation{Center for Spintronics Research Network, Tohoku University, Sendai 980-8578, Japan}

\author{Atsushi Fujimori}
\affiliation{Department of Physics, The University of Tokyo, Bunkyo-ku, Tokyo 113-0033, Japan}
\affiliation{Department of Applied Physics, Waseda University, Shinjuku-ku, Tokyo, 169-8555, Japan}



\date{\today}

\begin{abstract}
SmO thin film is a new Kondo system showing a resistivity upturn around 10 K and was theoretically proposed to have a topologically nontrivial band structure.
We have performed hard x-ray and soft x-ray photoemission spectroscopy to elucidate the electronic structure of SmO. From the Sm 3$d$ core-level spectra, we have estimated the valence of Sm to be $\sim$2.96, proving that the Sm has a mixed valence. The valence-band photoemission spectra exhibit a clear Fermi edge originating from the Sm 5$d$-derived band. The present finding is consistent with the theory suggesting a possible topological state in SmO and show that rare-earth monoxides or their heterostructures can be a new playground for the interplay of strong electron correlation and spin-orbit coupling.

\end{abstract}

\pacs{Valid PACS appear here}
\maketitle


\section{Introduction}

Rare-earth materials have been studied extensively for decades owing to their unique physical properties such as valence fluctuation \cite{Lawrence:1981aa}, Kondo insulating state \cite{Mason:1992aa}, heavy-fermion behavior \cite{Stewart:1984aa}, and superconductivity \cite{Mathur:1998aa}. Recently, a possible topological Kondo insulating state, where both strong spin-orbit coupling and electron-electron correlation play a role, has been proposed and attracted considerable attention \cite{Dzero:2010aa, Jiang:2013aa, Xu:2014aa}. 
Although the subjects of such studies were mostly limited to bulk crystals and their surfaces, the development of thin-film growth techniques has made it possible to grow single-crystalline samples even if the bulk crystal does not exist or is unstable.
Rare-earth monoxides (RO) (R: rare-earth element) are such examples, and they were successfully grown in a thin-film form by pulsed laser deposition (PLD) \cite{Kaminaga:2016aa,Uchida:2017aa,Kaminaga:2018aa,Kaminaga:2018ab,Saito:2019aa}, although RO quickly oxidizes in air to form R$_{2}$O$_{3}$. 
While R$_{2}$O$_{3}$ compounds are insulators with the localized $f$ electrons of the R$^{3+}$ ion, RO compounds have diverse and intriguing physical properties. YO seems to be a Mott insulator with the 4$d^{1}$ electronic configuration \cite{Kaminaga:2016aa}. LaO exhibits superconductivity \cite{Kaminaga:2018aa}. SmO \cite{Uchida:2017aa} shows dense Kondo effects. NdO is an itinerant ferromagnet \cite{Saito:2019aa}.

\begin{figure}
\begin{center}
\includegraphics[width=8.0cm]{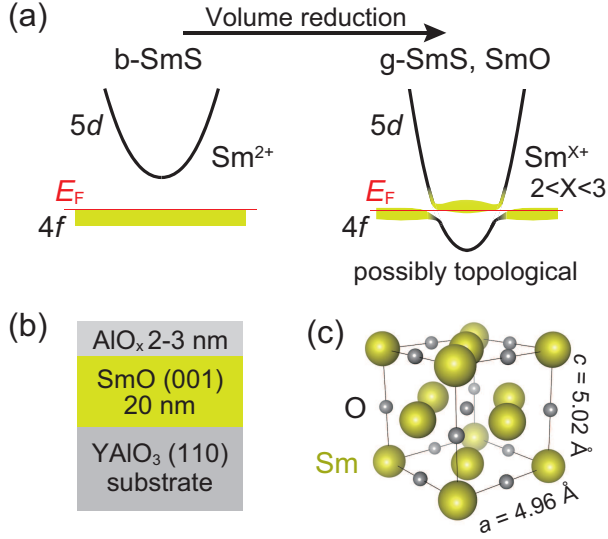}
\caption{(a) Schematic energy band structure of SmS and SmO. Reduction in the Sm-S or -O bond length leads to a mixed-valence state possibly with topologically nontrivial band structures. (b) Structure of thin-film samples. (c) Crystal structure of SmO thin films grown on YAlO$_{\rm 3}$ substrate.}
\label{Structure}
\end{center}
\end{figure}

Sm compounds are particularly interesting because some of them may have topologically nontrivial band structures, as suggested in refs \cite{Lu:2013aa,Kasinathan:2015aa,Li:2014aa,Kang:2015aa,Kang:2019aa}. 
SmS shows a first-order phase transition at the pressure of 6.5 kbar from the black phase (b-SmS), where Sm valence is 2+, to the golden phase (g-SmS), where Sm has a mixed valence of 2.6$\sim$2.8+ \cite{jayaraman:1970aa}. 
In the mixed-valence state, SmS is proposed to have a topologically nontrivial band structure \cite{Kang:2019aa}. 
This valence-state transition happens because the lattice-constant decrease lowers the bottom of the 5$d$ conduction band and the 4$f^{6}$5$d^{0}$ configuration, referred to as Sm$^{2+}$, starts to be mixed with the 4$f^{5}$5$d^{1}$ configuration, referred to as Sm$^{3+}$.
Such a volume-reduction induced mixed-valence phase would also appear if the anion size shrinks from S to O.
In fact, the SmO lattice constant of $\sim$5.0 \AA\ \cite{Uchida:2017aa} is smaller than the g-SmS and b-SmS lattice constants of $\sim$5.6 \AA\ and $\sim$5.9 \AA\ \cite{Deen:2005aa}. SmO was found to be metallic with resistivity upturn around 10 K followed by a decrease at even lower temperatures \cite{Uchida:2017aa} being similar to the case of g-SmS \cite{Imura:2009aa, Kang:2015aa}. 
SmO was also predicted to have a topological band structure \cite{Kasinathan:2015aa}. Such a scenario is illustrated schematically in Fig. \ref{Structure}(a).
As for the valence of Sm, which characterizes the electronic structure of 4$f$ electrons in SmO, Uchida $et$ $al$. \cite{Uchida:2017aa} estimated it to be 2.9+ by surface-sensitive soft x-ray (Al K$\alpha$) photoemission spectroscopy (SXPES). During the measurement, they etched the surface of SmO thin film by Ar-ion sputtering to overcome the surface sensitivity of SXPES. However, Ar-ion sputtering often degrades the surface and alters the chemical composition \cite{Steinberger:2015aa}. 

In the present study, to assess the valence of Sm and the electronic structure of SmO in a non-destructive way, we have performed hard x-ray photoemission spectroscopy (HAXPES) in addition to SXPES. HAXPES has a probing depth of 7-20 nm, about five times larger than that of SXPES \cite{Takata:2005aa}. Such an extended probing depth enabled us to study the bulk electronic structure through the capping layer without damaging samples. 
By decomposing the Sm 3$d$ HAXPES spectra into the Sm$^{3+}$ and Sm$^{2+}$ components, we have estimated the valence of Sm to be $\sim$2.96. We have observed a clear Fermi edge originating from the Sm 5$d$ band, consistent with the metallic conduction of the SmO thin films.

\section{Experiment \& Calculation}
SmO (001) epitaxial thin films were grown on YAlO$_{3}$ (110) substrates by PLD using a KrF excimer laser ($\lambda$ = 248 nm) with the energy output of 1.0 J/cm$^{2}$. Prior to growth, YAlO$_{3}$ substrates were annealed in a furnace at 1200°C for 4 hours to obtain an atomically flat surface. A Sm seed layer was grown at 400°C in vacuum (1.0 $\times$ 10$^{-8}$ Torr) at the pulse repetition rate of 1 Hz. Subsequently, a SmO film was grown on the seed layer in an Ar and O$_{2}$ mixed gas (Ar : O = 99 : 1, 5.0 $\times$ 10$^{-8}$ Torr in total) at the pulse repetition rate of 10 Hz. In order to protect the film from oxidation, an AlO$_{x}$ capping layer was grown $in-situ$ at room temperature. 
The sample structure was AlO$_{x}$ (2 or 3 nm)/SmO (20 nm)/YAlO$_{3}$(110) substrate (Fig. \ref{Structure}(b)), and the thickness of the capping layer was 3 nm for the HAXPES and 2 nm for the SXPES measurements.
The crystal structure of SmO was identified to be the pseudo-cubic rock-salt type with the lattice parameters of $a =  4.96$ \AA\ and $c = 5.02$ \AA\  by x-ray diffraction (Fig. \ref{Structure}(c)). 


HAXPES measurements were performed at the beamline BL09XU of SPring-8. The used photon energy was 8 keV, and the energy resolution was about 270 meV. The measurement temperature was 300 K.
SXPES measurements were performed at the beamline BL-2A of Photon Factory. The photon energy was varied between 1 keV to 1.3 keV to cover the Sm $M_{4,5}$ absorption edge. The typical energy resolution was about 250 meV. The measurement temperature was 11 K.

In order to simulate the experimental spectra, we have performed atomic multiplet calculations using a XTLS 8.5 code \cite{Tanaka:1994aa}. The ionic Hartree-Fock (HF) values  \cite{Yamasaki:2005aa} were used for the Slater integral, while the spin-orbit-interaction values were taken from refs. \cite{Utsumi:2017aa} \footnote{Slater integrals ($F$ and $G$) and spin orbit interaction ($\zeta$) used for Sm$^{3+}$ 3$d^{10}$4$f^{5}$ and 3$d^{9}$4$f^{5}$ are ($F^{2}_{f}$, $F^{4}_{f}$, $F^{6}_{f}$, $\zeta_{f}$) = (13.64, 8.56, 6.16, 0.144) and (15.31, 9.68, 6.98, 0.181), respectively, and  ($F^{2}_{fd}$, $F^{4}_{fd}$, $G^{1}_{fd}$, $G^{3}_{fd}$,$G^{5}_{fd}$,$\zeta_{d}$) = (9.63, 4.49, 6.87, 4.03, 2.78, 10.35)}. The calculated spectra were convolved with a Gaussian function with the full-width at half maxima (FWHM) of 0.27 eV to account for the energy resolution and also with a Lorentzian function with the FWHM of 1.2 eV to account for the core-hole lifetime broadening. We have also employed a Mahan function with the cut-off energy $\xi$ of 2 eV and the asymmetry parameter $\alpha$ of 0.7. 
These sets of parameters gave a reasonable agreement between the experiment and calculation, as will be shown below.

\section{Results and Discussion}

\begin{figure}
\begin{center}
\includegraphics[width=8.0cm]{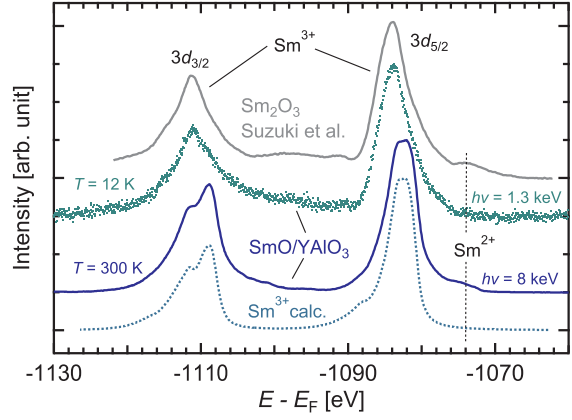}
\caption{Sm 3$d$ core-level photoemission spectra of SmO/YAlO$_{3}$ (blue solid curve) taken with hard x rays of $h\nu$ = 8 keV and soft x rays of 1.3 keV. Shirley backgrounds have been subtracted. The spectrum of Sm$_{2}$O$_{3}$ \cite{Suzuki:2000aa} is shown for reference. The result of atomic multiplet calculation is also shown by a dashed curve.}
\label{3d}
\end{center}
\end{figure}

Figure \ref{3d} shows the Sm 3$d$ core-level photoemission spectra of SmO thin films with Shirley-type backgrounds subtracted. The HAXPES spectrum of SmO/YAlO$_{3}$ is indicated by a blue solid curve, while the SXPES spectrum is shown by a green dotted curve. The spectral line shapes of the HAXPES and SXPES spectra are notably different from each other, accentuating the importance of the bulk sensitivity of HAXPES.
The SXPES spectrum looks similar to that of Sm$_{2}$O$_{3}$ \cite{Suzuki:2000aa}, probably because SmO oxidized near the surface. Note that SXPES is very surface sensitive for deep core levels like Sm 3$d$ as the kinetic energies of photoelectrons become small. 
The HAXPES spectrum for Sm 3$d_{5/2}$ consists of significant Sm$^{3+}$ signals around 1082 eV and a little Sm$^{2+}$ shoulder around 1074 eV, demonstrating that Sm is in a mixed-valence state. 
The line shape of the HAXPES spectrum of SmO is similar to that of g-SmS \cite{Mori:2007aa} and Sm metal \cite{Dallera:2005aa}, although there is a small difference in the Sm$^{3+}$ and Sm$^{2+}$ intensity ratios.
The Sm$^{3+}$ component of the HAXPES spectrum is reproduced by the atomic multiplet calculation reasonably well, as shown by a blue dashed curve in Fig. \ref{3d}. The agreement between theory and experiment and the fact that the HAXPES spectrum of SmO is markedly different from that of Sm$_{2}$O$_{3}$ indicate that the obtained HAXPES spectrum is intrinsic and reflects the electronic structure of SmO. 

\begin{figure}
\begin{center}
\includegraphics[width=7.8cm]{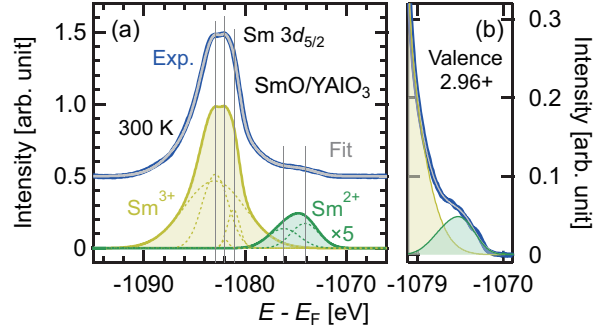}
\caption{(a) Sm 3$d_{5/2}$ core-level photoemission spectra of SmO/YAlO$_{3}$ (blue curve). The fitting results using multiple Gaussian functions are shown by gray curves. Each Gaussian function is shown by dashed curves, and the total Sm$^{3+}$ and Sm$^{2+}$ components are shown by yellow and green solid curves with shaded areas, respectively. (b) Magnified plot around the Sm$^{2+}$ regions.}
\label{ValenceEst}
\end{center}
\end{figure}

The Sm 3$d_{5/2}$ HAXPES spectrum is replotted in Fig. \ref{ValenceEst}(a) for the estimation of the valence of Sm. Since the atomic multiplet calculation did not perfectly match the experimental spectrum, we used multiple Gaussian functions to fit the spectral line shape and to decompose the spectrum into the Sm$^{3+}$ and Sm$^{2+}$ components.
Four Gaussian functions, which are shown by yellow dashed curves, were used to fit the Sm$^{3+}$ component and two Gaussian functions, shown by green dashed curves, were used to fit the the Sm$^{2+}$ component.
The summations of the Gaussian functions for each of the Sm$^{3+}$ and Sm$^{2+}$ components are indicated by yellow and green solid curves with shaded areas, respectively, and the entire fitting curve is shown by a gray solid curve, which is overlaid on the experimental spectrum (a blue solid curve). 
Such a fitting well reproduced the experimental spectrum not only in the Sm$^{3+}$ region but also in the Sm$^{2+}$ region, as can be seen in Fig. \ref{ValenceEst}(b), where the Sm$^{2+}$ shoulder is shown in detail.
The fitting yielded the Sm valence of $2.96 \pm 0.02$.

\begin{figure}
\begin{center}
\includegraphics[width=8.0cm]{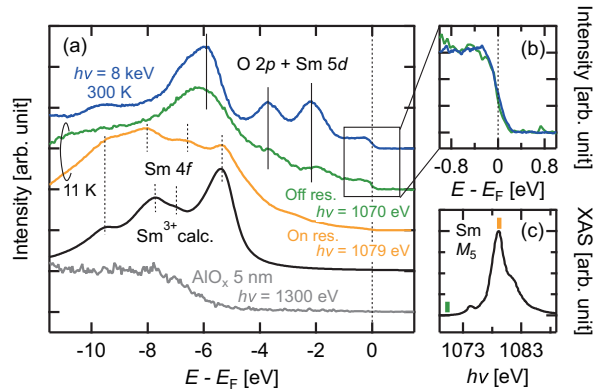}
\caption{(a) Valence-band photoemission spectra of SmO thin films. Soft x-rays around the Sm $M_{5}$ absorption edge were also used to enhance the 4$f$ contributions to the spectra. The atomic multiplet calculation for Sm$^{3+}$ 4$f$ is also shown. A sample with a thick AlO$_{x}$ capping layer was also measured and shown at the bottom. (b) Magnified plot around $E_{\rm F}$. (c) Sm $M_{5}$ x-ray absorption spectrum of SmO/YAlO$_{3}$.}
\label{Valence}
\end{center}
\end{figure}

Figure \ref{Valence}(a) shows the valence-band photoemission spectra of SmO/YAlO$_{3}$ taken with hard x rays of 8 keV and soft x rays of 1070 eV and 1079 eV. All the spectra have been normalized to their maxima. Since the signals from the capping layer, shown by a gray curve at the bottom, are featureless, especially near $E_{\rm F}$, the valence-band spectra obtained in the present study reflect the electronic structure of SmO.
Here, 1070 eV is slightly below the Sm $M_{5}$ absorption edge (off resonance), and 1079 eV is the energy exactly at the Sm $M_{5}$ absorption peak (on resonance). These photon energies are indicated by bars on the Sm $M_{5}$ x-ray absorption (XAS) spectrum shown in Fig. \ref{Valence}(c). There was a 120-fold photoemission intensity enhancement on the resonant condition.
The off-resonance spectrum looks qualitatively similar to the HAXPES spectrum. However, the on-resonance spectrum, which highlights the Sm 4$f$ contributions, looks different. The discrepancy between the HAXPES and the on-resonance spectra shows that the HAXPES spectrum having peaks at -2.2 eV, -3.7 eV, and -5.9 eV predominantly consist of the oxygen 2$p$ band strongly hybridized with Sm 5$d$ orbitals. 
Such a dominance of the O 2$p$ band in the HAXPES valence-band spectra rather than rare-earth 4$f$ peaks was also observed in Nd-containing oxides \cite{Eguchi:2009aa, Yamamoto:2018aa, Horio:2018aa}. 
The peak positions of the on-resonance spectra are reproduced by atomic multiplet calculation with the same parameters used for the core-level calculation, as shown by a black curve in Fig. \ref{Valence}(a).
Figure \ref{Valence}(b) shows a magnified view of the spectra near the Fermi level. A clear Fermi edge is observed, which most likely originates from the Sm 5$d$ bands as Sm atoms have the predominant Sm$^{3+}$ 4$f^{5}$5$d^{1}$ configuration. 

The present result that Sm has a mixed valence of 2.96+ having Sm 5$d$ bands crossing the Fermi level is consistent with the theory predicting a possible topological Kondo semi-metallic state in SmO \cite{Kasinathan:2015aa}. This new platform, namely epitaxial SmO thin film and its heterostructures with other RO compounds, would offer a variety of possibilities such as the strain-control of valence states and topological properties, quantum anomalous Hall effect if in contact with ferromagnetic EuO \cite{Kasinathan:2015aa}, and Majorana zero mode if in contact with superconducting LaO \cite{Lee:2014aa, Fu:2008aa}.


\section{Summary}
We have studied the electronic structure of the new Kondo system SmO thin films employing hard x-ray and soft x-ray photoemission spectroscopy. By decomposing the Sm 3$d$ core-level spectra into the Sm$^{3+}$ and Sm$^{2+}$ components, we have estimated the valence of Sm to be $\sim$2.96, proving that Sm indeed has a mixed valence. The valence-band photoemission spectra show that the Sm 5$d$ bands cross the Fermi level, consistent with the metallic conduction of the SmO thin films. The present findings are also compatible with the theory suggesting a possible topological state in SmO and show that the epitaxial SmO thin films and the heterostructures with other rare-earth monoxides can be a new playground for the interplay of strong electron correlation and spin-orbit coupling.

\section*{acknowledgments}
This work was supported by Grants-in-Aids for Scientific Research from JSPS (Grants No. 15H02109, No. 17H04922, No. 18H03872, No. 19K03741, and No. 19K15440).
The hard x-ray photoemission experiments were performed at beamline BL09XU of SPring-8 under the approval of  the Japan Synchrotron Radiation Research Institute (Proposal No. 2017A1113). The soft x-ray photoemission experiments were done at beamline BL2A under the approval of the Program Advisory Committee (Proposal No. 2016G096).
A.F. acknowledges support as an adjunct member of the Center for Spintronics Research Network (CSRN), the University of Tokyo, under Spintronics Research Network of Japan (Spin-RNJ).
S.S. and K. Kaminaga acknowledge financial support from the JSPS Research Fellowship for Young Scientists. 

\bibliography{../../Bibtex/BibTex_all}

\begin{thebibliography}{34}%
\makeatletter
\providecommand \@ifxundefined [1]{%
 \@ifx{#1\undefined}
}%
\providecommand \@ifnum [1]{%
 \ifnum #1\expandafter \@firstoftwo
 \else \expandafter \@secondoftwo
 \fi
}%
\providecommand \@ifx [1]{%
 \ifx #1\expandafter \@firstoftwo
 \else \expandafter \@secondoftwo
 \fi
}%
\providecommand \natexlab [1]{#1}%
\providecommand \enquote  [1]{``#1''}%
\providecommand \bibnamefont  [1]{#1}%
\providecommand \bibfnamefont [1]{#1}%
\providecommand \citenamefont [1]{#1}%
\providecommand \href@noop [0]{\@secondoftwo}%
\providecommand \href [0]{\begingroup \@sanitize@url \@href}%
\providecommand \@href[1]{\@@startlink{#1}\@@href}%
\providecommand \@@href[1]{\endgroup#1\@@endlink}%
\providecommand \@sanitize@url [0]{\catcode `\\12\catcode `\$12\catcode
  `\&12\catcode `\#12\catcode `\^12\catcode `\_12\catcode `\%12\relax}%
\providecommand \@@startlink[1]{}%
\providecommand \@@endlink[0]{}%
\providecommand \url  [0]{\begingroup\@sanitize@url \@url }%
\providecommand \@url [1]{\endgroup\@href {#1}{\urlprefix }}%
\providecommand \urlprefix  [0]{URL }%
\providecommand \Eprint [0]{\href }%
\providecommand \doibase [0]{http://dx.doi.org/}%
\providecommand \selectlanguage [0]{\@gobble}%
\providecommand \bibinfo  [0]{\@secondoftwo}%
\providecommand \bibfield  [0]{\@secondoftwo}%
\providecommand \translation [1]{[#1]}%
\providecommand \BibitemOpen [0]{}%
\providecommand \bibitemStop [0]{}%
\providecommand \bibitemNoStop [0]{.\EOS\space}%
\providecommand \EOS [0]{\spacefactor3000\relax}%
\providecommand \BibitemShut  [1]{\csname bibitem#1\endcsname}%
\let\auto@bib@innerbib\@empty
\bibitem [{\citenamefont {Lawrence}\ \emph {et~al.}(1981)\citenamefont
  {Lawrence}, \citenamefont {Riseborough},\ and\ \citenamefont
  {Parks}}]{Lawrence:1981aa}%
  \BibitemOpen
  \bibfield  {author} {\bibinfo {author} {\bibfnamefont {J.~M.}\ \bibnamefont
  {Lawrence}}, \bibinfo {author} {\bibfnamefont {P.~S.}\ \bibnamefont
  {Riseborough}}, \ and\ \bibinfo {author} {\bibfnamefont {R.~D.}\ \bibnamefont
  {Parks}},\ }\href {\doibase 10.1088/0034-4885/44/1/001} {\bibfield  {journal}
  {\bibinfo  {journal} {Rep. Prog. Phys.}\ }\textbf {\bibinfo {volume} {44}},\
  \bibinfo {pages} {1} (\bibinfo {year} {1981})}\BibitemShut {NoStop}%
\bibitem [{\citenamefont {Mason}\ \emph {et~al.}(1992)\citenamefont {Mason},
  \citenamefont {Aeppli}, \citenamefont {Ramirez}, \citenamefont {Clausen},
  \citenamefont {Broholm}, \citenamefont {St\"ucheli}, \citenamefont {Bucher},\
  and\ \citenamefont {Palstra}}]{Mason:1992aa}%
  \BibitemOpen
  \bibfield  {author} {\bibinfo {author} {\bibfnamefont {T.~E.}\ \bibnamefont
  {Mason}}, \bibinfo {author} {\bibfnamefont {G.}~\bibnamefont {Aeppli}},
  \bibinfo {author} {\bibfnamefont {A.~P.}\ \bibnamefont {Ramirez}}, \bibinfo
  {author} {\bibfnamefont {K.~N.}\ \bibnamefont {Clausen}}, \bibinfo {author}
  {\bibfnamefont {C.}~\bibnamefont {Broholm}}, \bibinfo {author} {\bibfnamefont
  {N.}~\bibnamefont {St\"ucheli}}, \bibinfo {author} {\bibfnamefont
  {E.}~\bibnamefont {Bucher}}, \ and\ \bibinfo {author} {\bibfnamefont
  {T.~T.~M.}\ \bibnamefont {Palstra}},\ }\href {\doibase
  10.1103/PhysRevLett.69.490} {\bibfield  {journal} {\bibinfo  {journal} {Phys.
  Rev. Lett.}\ }\textbf {\bibinfo {volume} {69}},\ \bibinfo {pages} {490}
  (\bibinfo {year} {1992})}\BibitemShut {NoStop}%
\bibitem [{\citenamefont {Stewart}(1984)}]{Stewart:1984aa}%
  \BibitemOpen
  \bibfield  {author} {\bibinfo {author} {\bibfnamefont {G.~R.}\ \bibnamefont
  {Stewart}},\ }\href {\doibase 10.1103/RevModPhys.56.755} {\bibfield
  {journal} {\bibinfo  {journal} {Rev. Mod. Phys.}\ }\textbf {\bibinfo {volume}
  {56}},\ \bibinfo {pages} {755} (\bibinfo {year} {1984})}\BibitemShut
  {NoStop}%
\bibitem [{\citenamefont {Mathur}\ \emph {et~al.}(1998)\citenamefont {Mathur},
  \citenamefont {Grosche}, \citenamefont {Julian}, \citenamefont {Walker},
  \citenamefont {Freye}, \citenamefont {Haselwimmer},\ and\ \citenamefont
  {Lonzarich}}]{Mathur:1998aa}%
  \BibitemOpen
  \bibfield  {author} {\bibinfo {author} {\bibfnamefont {N.}~\bibnamefont
  {Mathur}}, \bibinfo {author} {\bibfnamefont {F.}~\bibnamefont {Grosche}},
  \bibinfo {author} {\bibfnamefont {S.}~\bibnamefont {Julian}}, \bibinfo
  {author} {\bibfnamefont {I.}~\bibnamefont {Walker}}, \bibinfo {author}
  {\bibfnamefont {D.}~\bibnamefont {Freye}}, \bibinfo {author} {\bibfnamefont
  {R.}~\bibnamefont {Haselwimmer}}, \ and\ \bibinfo {author} {\bibfnamefont
  {G.}~\bibnamefont {Lonzarich}},\ }\href@noop {} {\bibfield  {journal}
  {\bibinfo  {journal} {Nature}\ }\textbf {\bibinfo {volume} {394}},\ \bibinfo
  {pages} {39} (\bibinfo {year} {1998})}\BibitemShut {NoStop}%
\bibitem [{\citenamefont {Dzero}\ \emph {et~al.}(2010)\citenamefont {Dzero},
  \citenamefont {Sun}, \citenamefont {Galitski},\ and\ \citenamefont
  {Coleman}}]{Dzero:2010aa}%
  \BibitemOpen
  \bibfield  {author} {\bibinfo {author} {\bibfnamefont {M.}~\bibnamefont
  {Dzero}}, \bibinfo {author} {\bibfnamefont {K.}~\bibnamefont {Sun}}, \bibinfo
  {author} {\bibfnamefont {V.}~\bibnamefont {Galitski}}, \ and\ \bibinfo
  {author} {\bibfnamefont {P.}~\bibnamefont {Coleman}},\ }\href {\doibase
  10.1103/PhysRevLett.104.106408} {\bibfield  {journal} {\bibinfo  {journal}
  {Phys. Rev. Lett.}\ }\textbf {\bibinfo {volume} {104}},\ \bibinfo {pages}
  {106408} (\bibinfo {year} {2010})}\BibitemShut {NoStop}%
\bibitem [{\citenamefont {Jiang}\ \emph {et~al.}(2013)\citenamefont {Jiang},
  \citenamefont {Li}, \citenamefont {Zhang}, \citenamefont {Sun}, \citenamefont
  {Chen}, \citenamefont {Ye}, \citenamefont {Xu}, \citenamefont {Ge},
  \citenamefont {Tan}, \citenamefont {Niu}, \citenamefont {Xia}, \citenamefont
  {Xie}, \citenamefont {Li}, \citenamefont {Chen}, \citenamefont {Wen},\ and\
  \citenamefont {Feng}}]{Jiang:2013aa}%
  \BibitemOpen
  \bibfield  {author} {\bibinfo {author} {\bibfnamefont {J.}~\bibnamefont
  {Jiang}}, \bibinfo {author} {\bibfnamefont {S.}~\bibnamefont {Li}}, \bibinfo
  {author} {\bibfnamefont {T.}~\bibnamefont {Zhang}}, \bibinfo {author}
  {\bibfnamefont {Z.}~\bibnamefont {Sun}}, \bibinfo {author} {\bibfnamefont
  {F.}~\bibnamefont {Chen}}, \bibinfo {author} {\bibfnamefont {Z.~R.}\
  \bibnamefont {Ye}}, \bibinfo {author} {\bibfnamefont {M.}~\bibnamefont {Xu}},
  \bibinfo {author} {\bibfnamefont {Q.~Q.}\ \bibnamefont {Ge}}, \bibinfo
  {author} {\bibfnamefont {S.~Y.}\ \bibnamefont {Tan}}, \bibinfo {author}
  {\bibfnamefont {X.~H.}\ \bibnamefont {Niu}}, \bibinfo {author} {\bibfnamefont
  {M.}~\bibnamefont {Xia}}, \bibinfo {author} {\bibfnamefont {B.~P.}\
  \bibnamefont {Xie}}, \bibinfo {author} {\bibfnamefont {Y.~F.}\ \bibnamefont
  {Li}}, \bibinfo {author} {\bibfnamefont {X.~H.}\ \bibnamefont {Chen}},
  \bibinfo {author} {\bibfnamefont {H.~H.}\ \bibnamefont {Wen}}, \ and\
  \bibinfo {author} {\bibfnamefont {D.~L.}\ \bibnamefont {Feng}},\ }\href@noop
  {} {\bibfield  {journal} {\bibinfo  {journal} {Nat. Commun.}\ }\textbf
  {\bibinfo {volume} {4}},\ \bibinfo {pages} {3010} (\bibinfo {year}
  {2013})}\BibitemShut {NoStop}%
\bibitem [{\citenamefont {Xu}\ \emph {et~al.}(2014)\citenamefont {Xu},
  \citenamefont {Biswas}, \citenamefont {Dil}, \citenamefont {Dhaka},
  \citenamefont {Landolt}, \citenamefont {Muff}, \citenamefont {Matt},
  \citenamefont {Shi}, \citenamefont {Plumb}, \citenamefont {Radovi{\'c}},
  \citenamefont {Pomjakushina}, \citenamefont {Conder}, \citenamefont {Amato},
  \citenamefont {Borisenko}, \citenamefont {Yu}, \citenamefont {Weng},
  \citenamefont {Fang}, \citenamefont {Dai}, \citenamefont {Mesot},
  \citenamefont {Ding},\ and\ \citenamefont {Shi}}]{Xu:2014aa}%
  \BibitemOpen
  \bibfield  {author} {\bibinfo {author} {\bibfnamefont {N.}~\bibnamefont
  {Xu}}, \bibinfo {author} {\bibfnamefont {P.~K.}\ \bibnamefont {Biswas}},
  \bibinfo {author} {\bibfnamefont {J.~H.}\ \bibnamefont {Dil}}, \bibinfo
  {author} {\bibfnamefont {R.~S.}\ \bibnamefont {Dhaka}}, \bibinfo {author}
  {\bibfnamefont {G.}~\bibnamefont {Landolt}}, \bibinfo {author} {\bibfnamefont
  {S.}~\bibnamefont {Muff}}, \bibinfo {author} {\bibfnamefont {C.~E.}\
  \bibnamefont {Matt}}, \bibinfo {author} {\bibfnamefont {X.}~\bibnamefont
  {Shi}}, \bibinfo {author} {\bibfnamefont {N.~C.}\ \bibnamefont {Plumb}},
  \bibinfo {author} {\bibfnamefont {M.}~\bibnamefont {Radovi{\'c}}}, \bibinfo
  {author} {\bibfnamefont {E.}~\bibnamefont {Pomjakushina}}, \bibinfo {author}
  {\bibfnamefont {K.}~\bibnamefont {Conder}}, \bibinfo {author} {\bibfnamefont
  {A.}~\bibnamefont {Amato}}, \bibinfo {author} {\bibfnamefont {S.~V.}\
  \bibnamefont {Borisenko}}, \bibinfo {author} {\bibfnamefont {R.}~\bibnamefont
  {Yu}}, \bibinfo {author} {\bibfnamefont {H.-M.}\ \bibnamefont {Weng}},
  \bibinfo {author} {\bibfnamefont {Z.}~\bibnamefont {Fang}}, \bibinfo {author}
  {\bibfnamefont {X.}~\bibnamefont {Dai}}, \bibinfo {author} {\bibfnamefont
  {J.}~\bibnamefont {Mesot}}, \bibinfo {author} {\bibfnamefont
  {H.}~\bibnamefont {Ding}}, \ and\ \bibinfo {author} {\bibfnamefont
  {M.}~\bibnamefont {Shi}},\ }\href@noop {} {\bibfield  {journal} {\bibinfo
  {journal} {Nat. Commun.}\ }\textbf {\bibinfo {volume} {5}},\ \bibinfo {pages}
  {4566} (\bibinfo {year} {2014})}\BibitemShut {NoStop}%
\bibitem [{\citenamefont {Kaminaga}\ \emph {et~al.}(2016)\citenamefont
  {Kaminaga}, \citenamefont {Sei}, \citenamefont {Hayashi}, \citenamefont
  {Happo}, \citenamefont {Tajiri}, \citenamefont {Oka}, \citenamefont
  {Fukumura},\ and\ \citenamefont {Hasegawa}}]{Kaminaga:2016aa}%
  \BibitemOpen
  \bibfield  {author} {\bibinfo {author} {\bibfnamefont {K.}~\bibnamefont
  {Kaminaga}}, \bibinfo {author} {\bibfnamefont {R.}~\bibnamefont {Sei}},
  \bibinfo {author} {\bibfnamefont {K.}~\bibnamefont {Hayashi}}, \bibinfo
  {author} {\bibfnamefont {N.}~\bibnamefont {Happo}}, \bibinfo {author}
  {\bibfnamefont {H.}~\bibnamefont {Tajiri}}, \bibinfo {author} {\bibfnamefont
  {D.}~\bibnamefont {Oka}}, \bibinfo {author} {\bibfnamefont {T.}~\bibnamefont
  {Fukumura}}, \ and\ \bibinfo {author} {\bibfnamefont {T.}~\bibnamefont
  {Hasegawa}},\ }\href {\doibase 10.1063/1.4944330} {\bibfield  {journal}
  {\bibinfo  {journal} {Appl. Phys. Lett.}\ }\textbf {\bibinfo {volume}
  {108}},\ \bibinfo {pages} {122102} (\bibinfo {year} {2016})}\BibitemShut
  {NoStop}%
\bibitem [{\citenamefont {Uchida}\ \emph {et~al.}(2017)\citenamefont {Uchida},
  \citenamefont {Kaminaga}, \citenamefont {Fukumura},\ and\ \citenamefont
  {Hasegawa}}]{Uchida:2017aa}%
  \BibitemOpen
  \bibfield  {author} {\bibinfo {author} {\bibfnamefont {Y.}~\bibnamefont
  {Uchida}}, \bibinfo {author} {\bibfnamefont {K.}~\bibnamefont {Kaminaga}},
  \bibinfo {author} {\bibfnamefont {T.}~\bibnamefont {Fukumura}}, \ and\
  \bibinfo {author} {\bibfnamefont {T.}~\bibnamefont {Hasegawa}},\ }\href
  {\doibase 10.1103/PhysRevB.95.125111} {\bibfield  {journal} {\bibinfo
  {journal} {Phys. Rev. B}\ }\textbf {\bibinfo {volume} {95}},\ \bibinfo
  {pages} {125111} (\bibinfo {year} {2017})}\BibitemShut {NoStop}%
\bibitem [{\citenamefont {Kaminaga}\ \emph
  {et~al.}(2018{\natexlab{a}})\citenamefont {Kaminaga}, \citenamefont {Oka},
  \citenamefont {Hasegawa},\ and\ \citenamefont {Fukumura}}]{Kaminaga:2018aa}%
  \BibitemOpen
  \bibfield  {author} {\bibinfo {author} {\bibfnamefont {K.}~\bibnamefont
  {Kaminaga}}, \bibinfo {author} {\bibfnamefont {D.}~\bibnamefont {Oka}},
  \bibinfo {author} {\bibfnamefont {T.}~\bibnamefont {Hasegawa}}, \ and\
  \bibinfo {author} {\bibfnamefont {T.}~\bibnamefont {Fukumura}},\ }\href
  {\doibase 10.1021/jacs.8b03009} {\bibfield  {journal} {\bibinfo  {journal}
  {J. Am. Chem. Soc.}\ }\textbf {\bibinfo {volume} {140}},\ \bibinfo {pages}
  {6754} (\bibinfo {year} {2018}{\natexlab{a}})}\BibitemShut {NoStop}%
\bibitem [{\citenamefont {Kaminaga}\ \emph
  {et~al.}(2018{\natexlab{b}})\citenamefont {Kaminaga}, \citenamefont {Oka},
  \citenamefont {Hasegawa},\ and\ \citenamefont {Fukumura}}]{Kaminaga:2018ab}%
  \BibitemOpen
  \bibfield  {author} {\bibinfo {author} {\bibfnamefont {K.}~\bibnamefont
  {Kaminaga}}, \bibinfo {author} {\bibfnamefont {D.}~\bibnamefont {Oka}},
  \bibinfo {author} {\bibfnamefont {T.}~\bibnamefont {Hasegawa}}, \ and\
  \bibinfo {author} {\bibfnamefont {T.}~\bibnamefont {Fukumura}},\ }\href
  {\doibase 10.1021/acsomega.8b02082} {\bibfield  {journal} {\bibinfo
  {journal} {ACS Omega}\ }\textbf {\bibinfo {volume} {3}},\ \bibinfo {pages}
  {12501} (\bibinfo {year} {2018}{\natexlab{b}})}\BibitemShut {NoStop}%
\bibitem [{\citenamefont {Saito}\ \emph {et~al.}(2019)\citenamefont {Saito},
  \citenamefont {Kaminaga}, \citenamefont {Oka},\ and\ \citenamefont
  {Fukumura}}]{Saito:2019aa}%
  \BibitemOpen
  \bibfield  {author} {\bibinfo {author} {\bibfnamefont {D.}~\bibnamefont
  {Saito}}, \bibinfo {author} {\bibfnamefont {K.}~\bibnamefont {Kaminaga}},
  \bibinfo {author} {\bibfnamefont {D.}~\bibnamefont {Oka}}, \ and\ \bibinfo
  {author} {\bibfnamefont {T.}~\bibnamefont {Fukumura}},\ }\href {\doibase
  10.1103/PhysRevMaterials.3.064407} {\bibfield  {journal} {\bibinfo  {journal}
  {Phys. Rev. Mater.}\ }\textbf {\bibinfo {volume} {3}},\ \bibinfo {pages}
  {064407} (\bibinfo {year} {2019})}\BibitemShut {NoStop}%
\bibitem [{\citenamefont {Lu}\ \emph {et~al.}(2013)\citenamefont {Lu},
  \citenamefont {Zhao}, \citenamefont {Weng}, \citenamefont {Fang},\ and\
  \citenamefont {Dai}}]{Lu:2013aa}%
  \BibitemOpen
  \bibfield  {author} {\bibinfo {author} {\bibfnamefont {F.}~\bibnamefont
  {Lu}}, \bibinfo {author} {\bibfnamefont {J.}~\bibnamefont {Zhao}}, \bibinfo
  {author} {\bibfnamefont {H.}~\bibnamefont {Weng}}, \bibinfo {author}
  {\bibfnamefont {Z.}~\bibnamefont {Fang}}, \ and\ \bibinfo {author}
  {\bibfnamefont {X.}~\bibnamefont {Dai}},\ }\href {\doibase
  10.1103/PhysRevLett.110.096401} {\bibfield  {journal} {\bibinfo  {journal}
  {Phys. Rev. Lett.}\ }\textbf {\bibinfo {volume} {110}},\ \bibinfo {pages}
  {096401} (\bibinfo {year} {2013})}\BibitemShut {NoStop}%
\bibitem [{\citenamefont {Kasinathan}\ \emph {et~al.}(2015)\citenamefont
  {Kasinathan}, \citenamefont {Koepernik}, \citenamefont {Tjeng},\ and\
  \citenamefont {Haverkort}}]{Kasinathan:2015aa}%
  \BibitemOpen
  \bibfield  {author} {\bibinfo {author} {\bibfnamefont {D.}~\bibnamefont
  {Kasinathan}}, \bibinfo {author} {\bibfnamefont {K.}~\bibnamefont
  {Koepernik}}, \bibinfo {author} {\bibfnamefont {L.~H.}\ \bibnamefont
  {Tjeng}}, \ and\ \bibinfo {author} {\bibfnamefont {M.~W.}\ \bibnamefont
  {Haverkort}},\ }\href {\doibase 10.1103/PhysRevB.91.195127} {\bibfield
  {journal} {\bibinfo  {journal} {Phys. Rev. B}\ }\textbf {\bibinfo {volume}
  {91}},\ \bibinfo {pages} {195127} (\bibinfo {year} {2015})}\BibitemShut
  {NoStop}%
\bibitem [{\citenamefont {Li}\ \emph {et~al.}(2014)\citenamefont {Li},
  \citenamefont {Li}, \citenamefont {Blaha},\ and\ \citenamefont
  {Kioussis}}]{Li:2014aa}%
  \BibitemOpen
  \bibfield  {author} {\bibinfo {author} {\bibfnamefont {Z.}~\bibnamefont
  {Li}}, \bibinfo {author} {\bibfnamefont {J.}~\bibnamefont {Li}}, \bibinfo
  {author} {\bibfnamefont {P.}~\bibnamefont {Blaha}}, \ and\ \bibinfo {author}
  {\bibfnamefont {N.}~\bibnamefont {Kioussis}},\ }\href {\doibase
  10.1103/PhysRevB.89.121117} {\bibfield  {journal} {\bibinfo  {journal} {Phys.
  Rev. B}\ }\textbf {\bibinfo {volume} {89}},\ \bibinfo {pages} {121117}
  (\bibinfo {year} {2014})}\BibitemShut {NoStop}%
\bibitem [{\citenamefont {Kang}\ \emph {et~al.}(2015)\citenamefont {Kang},
  \citenamefont {Choi}, \citenamefont {Kim},\ and\ \citenamefont
  {Min}}]{Kang:2015aa}%
  \BibitemOpen
  \bibfield  {author} {\bibinfo {author} {\bibfnamefont {C.-J.}\ \bibnamefont
  {Kang}}, \bibinfo {author} {\bibfnamefont {H.~C.}\ \bibnamefont {Choi}},
  \bibinfo {author} {\bibfnamefont {K.}~\bibnamefont {Kim}}, \ and\ \bibinfo
  {author} {\bibfnamefont {B.~I.}\ \bibnamefont {Min}},\ }\href {\doibase
  10.1103/PhysRevLett.114.166404} {\bibfield  {journal} {\bibinfo  {journal}
  {Phys. Rev. Lett.}\ }\textbf {\bibinfo {volume} {114}},\ \bibinfo {pages}
  {166404} (\bibinfo {year} {2015})}\BibitemShut {NoStop}%
\bibitem [{\citenamefont {Kang}\ \emph {et~al.}(2019)\citenamefont {Kang},
  \citenamefont {Ryu}, \citenamefont {Kim}, \citenamefont {Kim}, \citenamefont
  {Kang}, \citenamefont {Denlinger}, \citenamefont {Kotliar},\ and\
  \citenamefont {Min}}]{Kang:2019aa}%
  \BibitemOpen
  \bibfield  {author} {\bibinfo {author} {\bibfnamefont {C.-J.}\ \bibnamefont
  {Kang}}, \bibinfo {author} {\bibfnamefont {D.-C.}\ \bibnamefont {Ryu}},
  \bibinfo {author} {\bibfnamefont {J.}~\bibnamefont {Kim}}, \bibinfo {author}
  {\bibfnamefont {K.}~\bibnamefont {Kim}}, \bibinfo {author} {\bibfnamefont
  {J.-S.}\ \bibnamefont {Kang}}, \bibinfo {author} {\bibfnamefont {J.~D.}\
  \bibnamefont {Denlinger}}, \bibinfo {author} {\bibfnamefont {G.}~\bibnamefont
  {Kotliar}}, \ and\ \bibinfo {author} {\bibfnamefont {B.~I.}\ \bibnamefont
  {Min}},\ }\href {\doibase 10.1103/PhysRevMaterials.3.081201} {\bibfield
  {journal} {\bibinfo  {journal} {Phys. Rev. Mater.}\ }\textbf {\bibinfo
  {volume} {3}},\ \bibinfo {pages} {081201} (\bibinfo {year}
  {2019})}\BibitemShut {NoStop}%
\bibitem [{\citenamefont {Jayaraman}\ \emph {et~al.}(1970)\citenamefont
  {Jayaraman}, \citenamefont {Narayanamurti}, \citenamefont {Bucher},\ and\
  \citenamefont {Maines}}]{jayaraman:1970aa}%
  \BibitemOpen
  \bibfield  {author} {\bibinfo {author} {\bibfnamefont {A.}~\bibnamefont
  {Jayaraman}}, \bibinfo {author} {\bibfnamefont {V.}~\bibnamefont
  {Narayanamurti}}, \bibinfo {author} {\bibfnamefont {E.}~\bibnamefont
  {Bucher}}, \ and\ \bibinfo {author} {\bibfnamefont {R.~G.}\ \bibnamefont
  {Maines}},\ }\href {\doibase 10.1103/PhysRevLett.25.1430} {\bibfield
  {journal} {\bibinfo  {journal} {Phys. Rev. Lett.}\ }\textbf {\bibinfo
  {volume} {25}},\ \bibinfo {pages} {1430} (\bibinfo {year}
  {1970})}\BibitemShut {NoStop}%
\bibitem [{\citenamefont {Deen}\ \emph {et~al.}(2005)\citenamefont {Deen},
  \citenamefont {Braithwaite}, \citenamefont {Kernavanois}, \citenamefont
  {Paolasini}, \citenamefont {Raymond}, \citenamefont {Barla}, \citenamefont
  {Lapertot},\ and\ \citenamefont {Sanchez}}]{Deen:2005aa}%
  \BibitemOpen
  \bibfield  {author} {\bibinfo {author} {\bibfnamefont {P.~P.}\ \bibnamefont
  {Deen}}, \bibinfo {author} {\bibfnamefont {D.}~\bibnamefont {Braithwaite}},
  \bibinfo {author} {\bibfnamefont {N.}~\bibnamefont {Kernavanois}}, \bibinfo
  {author} {\bibfnamefont {L.}~\bibnamefont {Paolasini}}, \bibinfo {author}
  {\bibfnamefont {S.}~\bibnamefont {Raymond}}, \bibinfo {author} {\bibfnamefont
  {A.}~\bibnamefont {Barla}}, \bibinfo {author} {\bibfnamefont
  {G.}~\bibnamefont {Lapertot}}, \ and\ \bibinfo {author} {\bibfnamefont
  {J.~P.}\ \bibnamefont {Sanchez}},\ }\href {\doibase
  10.1103/PhysRevB.71.245118} {\bibfield  {journal} {\bibinfo  {journal} {Phys.
  Rev. B}\ }\textbf {\bibinfo {volume} {71}},\ \bibinfo {pages} {245118}
  (\bibinfo {year} {2005})}\BibitemShut {NoStop}%
\bibitem [{\citenamefont {Imura}\ \emph {et~al.}(2009)\citenamefont {Imura},
  \citenamefont {Matsubayashi}, \citenamefont {Suzuki}, \citenamefont
  {Deguchi},\ and\ \citenamefont {Sato}}]{Imura:2009aa}%
  \BibitemOpen
  \bibfield  {author} {\bibinfo {author} {\bibfnamefont {K.}~\bibnamefont
  {Imura}}, \bibinfo {author} {\bibfnamefont {K.}~\bibnamefont {Matsubayashi}},
  \bibinfo {author} {\bibfnamefont {H.}~\bibnamefont {Suzuki}}, \bibinfo
  {author} {\bibfnamefont {K.}~\bibnamefont {Deguchi}}, \ and\ \bibinfo
  {author} {\bibfnamefont {N.}~\bibnamefont {Sato}},\ }\href {\doibase
  https://doi.org/10.1016/j.physb.2009.07.013} {\bibfield  {journal} {\bibinfo
  {journal} {Physica B: Condensed Matter}\ }\textbf {\bibinfo {volume} {404}},\
  \bibinfo {pages} {3028 } (\bibinfo {year} {2009})}\BibitemShut {NoStop}%
\bibitem [{\citenamefont {Steinberger}\ \emph {et~al.}(2015)\citenamefont
  {Steinberger}, \citenamefont {Walter}, \citenamefont {Greunz}, \citenamefont
  {Duchoslav}, \citenamefont {Arndt}, \citenamefont {Molodtsov}, \citenamefont
  {Meyer},\ and\ \citenamefont {Stifter}}]{Steinberger:2015aa}%
  \BibitemOpen
  \bibfield  {author} {\bibinfo {author} {\bibfnamefont {R.}~\bibnamefont
  {Steinberger}}, \bibinfo {author} {\bibfnamefont {J.}~\bibnamefont {Walter}},
  \bibinfo {author} {\bibfnamefont {T.}~\bibnamefont {Greunz}}, \bibinfo
  {author} {\bibfnamefont {J.}~\bibnamefont {Duchoslav}}, \bibinfo {author}
  {\bibfnamefont {M.}~\bibnamefont {Arndt}}, \bibinfo {author} {\bibfnamefont
  {S.}~\bibnamefont {Molodtsov}}, \bibinfo {author} {\bibfnamefont
  {D.}~\bibnamefont {Meyer}}, \ and\ \bibinfo {author} {\bibfnamefont
  {D.}~\bibnamefont {Stifter}},\ }\href {\doibase
  https://doi.org/10.1016/j.corsci.2015.06.019} {\bibfield  {journal} {\bibinfo
   {journal} {Corros. Sci.}\ }\textbf {\bibinfo {volume} {99}},\ \bibinfo
  {pages} {66 } (\bibinfo {year} {2015})}\BibitemShut {NoStop}%
\bibitem [{\citenamefont {Takata}\ \emph {et~al.}(2005)\citenamefont {Takata},
  \citenamefont {Yabashi}, \citenamefont {Tamasaku}, \citenamefont {Nishino},
  \citenamefont {Miwa}, \citenamefont {Ishikawa}, \citenamefont {Ikenaga},
  \citenamefont {Horiba}, \citenamefont {Shin}, \citenamefont {Arita},
  \citenamefont {Shimada}, \citenamefont {Namatame}, \citenamefont {Taniguchi},
  \citenamefont {Nohira}, \citenamefont {Hattori}, \citenamefont
  {S{\"o}dergren}, \citenamefont {Wannberg},\ and\ \citenamefont
  {Kobayashi}}]{Takata:2005aa}%
  \BibitemOpen
  \bibfield  {author} {\bibinfo {author} {\bibfnamefont {Y.}~\bibnamefont
  {Takata}}, \bibinfo {author} {\bibfnamefont {M.}~\bibnamefont {Yabashi}},
  \bibinfo {author} {\bibfnamefont {K.}~\bibnamefont {Tamasaku}}, \bibinfo
  {author} {\bibfnamefont {Y.}~\bibnamefont {Nishino}}, \bibinfo {author}
  {\bibfnamefont {D.}~\bibnamefont {Miwa}}, \bibinfo {author} {\bibfnamefont
  {T.}~\bibnamefont {Ishikawa}}, \bibinfo {author} {\bibfnamefont
  {E.}~\bibnamefont {Ikenaga}}, \bibinfo {author} {\bibfnamefont
  {K.}~\bibnamefont {Horiba}}, \bibinfo {author} {\bibfnamefont
  {S.}~\bibnamefont {Shin}}, \bibinfo {author} {\bibfnamefont {M.}~\bibnamefont
  {Arita}}, \bibinfo {author} {\bibfnamefont {K.}~\bibnamefont {Shimada}},
  \bibinfo {author} {\bibfnamefont {H.}~\bibnamefont {Namatame}}, \bibinfo
  {author} {\bibfnamefont {M.}~\bibnamefont {Taniguchi}}, \bibinfo {author}
  {\bibfnamefont {H.}~\bibnamefont {Nohira}}, \bibinfo {author} {\bibfnamefont
  {T.}~\bibnamefont {Hattori}}, \bibinfo {author} {\bibfnamefont
  {S.}~\bibnamefont {S{\"o}dergren}}, \bibinfo {author} {\bibfnamefont
  {B.}~\bibnamefont {Wannberg}}, \ and\ \bibinfo {author} {\bibfnamefont
  {K.}~\bibnamefont {Kobayashi}},\ }\href {\doibase
  https://doi.org/10.1016/j.nima.2005.05.011} {\bibfield  {journal} {\bibinfo
  {journal} {Nucl. Instrum. Methods Phys. Res. A}\ }\textbf {\bibinfo {volume}
  {547}},\ \bibinfo {pages} {50 } (\bibinfo {year} {2005})},\ \bibinfo {note}
  {proceedings of the Workshop on Hard X-ray Photoelectron
  Spectroscopy}\BibitemShut {NoStop}%
\bibitem [{\citenamefont {Tanaka}\ and\ \citenamefont
  {Jo}(1994)}]{Tanaka:1994aa}%
  \BibitemOpen
  \bibfield  {author} {\bibinfo {author} {\bibfnamefont {A.}~\bibnamefont
  {Tanaka}}\ and\ \bibinfo {author} {\bibfnamefont {T.}~\bibnamefont {Jo}},\
  }\href {\doibase 10.1143/JPSJ.63.2788} {\bibfield  {journal} {\bibinfo
  {journal} {J. Phys. Soc. Jpn}\ }\textbf {\bibinfo {volume} {63}},\ \bibinfo
  {pages} {2788} (\bibinfo {year} {1994})}\BibitemShut {NoStop}%
\bibitem [{\citenamefont {Yamasaki}\ \emph {et~al.}(2005)\citenamefont
  {Yamasaki}, \citenamefont {Sekiyama}, \citenamefont {Imada}, \citenamefont
  {Tsunekawa}, \citenamefont {Dallera}, \citenamefont {Braicovich},
  \citenamefont {Lee}, \citenamefont {Ochiai},\ and\ \citenamefont
  {Suga}}]{Yamasaki:2005aa}%
  \BibitemOpen
  \bibfield  {author} {\bibinfo {author} {\bibfnamefont {A.}~\bibnamefont
  {Yamasaki}}, \bibinfo {author} {\bibfnamefont {A.}~\bibnamefont {Sekiyama}},
  \bibinfo {author} {\bibfnamefont {S.}~\bibnamefont {Imada}}, \bibinfo
  {author} {\bibfnamefont {M.}~\bibnamefont {Tsunekawa}}, \bibinfo {author}
  {\bibfnamefont {C.}~\bibnamefont {Dallera}}, \bibinfo {author} {\bibfnamefont
  {L.}~\bibnamefont {Braicovich}}, \bibinfo {author} {\bibfnamefont {T.-L.}\
  \bibnamefont {Lee}}, \bibinfo {author} {\bibfnamefont {A.}~\bibnamefont
  {Ochiai}}, \ and\ \bibinfo {author} {\bibfnamefont {S.}~\bibnamefont
  {Suga}},\ }\href {\doibase 10.1143/jpsj.74.2538} {\bibfield  {journal}
  {\bibinfo  {journal} {J. Phys. Soc. Jpn}\ }\textbf {\bibinfo {volume} {74}},\
  \bibinfo {pages} {2538} (\bibinfo {year} {2005})}\BibitemShut {NoStop}%
\bibitem [{\citenamefont {Utsumi}\ \emph {et~al.}(2017)\citenamefont {Utsumi},
  \citenamefont {Kasinathan}, \citenamefont {Ko}, \citenamefont {Agrestini},
  \citenamefont {Haverkort}, \citenamefont {Wirth}, \citenamefont {Wu},
  \citenamefont {Tsuei}, \citenamefont {Kim}, \citenamefont {Fisk},
  \citenamefont {Tanaka}, \citenamefont {Thalmeier},\ and\ \citenamefont
  {Tjeng}}]{Utsumi:2017aa}%
  \BibitemOpen
  \bibfield  {author} {\bibinfo {author} {\bibfnamefont {Y.}~\bibnamefont
  {Utsumi}}, \bibinfo {author} {\bibfnamefont {D.}~\bibnamefont {Kasinathan}},
  \bibinfo {author} {\bibfnamefont {K.-T.}\ \bibnamefont {Ko}}, \bibinfo
  {author} {\bibfnamefont {S.}~\bibnamefont {Agrestini}}, \bibinfo {author}
  {\bibfnamefont {M.~W.}\ \bibnamefont {Haverkort}}, \bibinfo {author}
  {\bibfnamefont {S.}~\bibnamefont {Wirth}}, \bibinfo {author} {\bibfnamefont
  {Y.-H.}\ \bibnamefont {Wu}}, \bibinfo {author} {\bibfnamefont {K.-D.}\
  \bibnamefont {Tsuei}}, \bibinfo {author} {\bibfnamefont {D.-J.}\ \bibnamefont
  {Kim}}, \bibinfo {author} {\bibfnamefont {Z.}~\bibnamefont {Fisk}}, \bibinfo
  {author} {\bibfnamefont {A.}~\bibnamefont {Tanaka}}, \bibinfo {author}
  {\bibfnamefont {P.}~\bibnamefont {Thalmeier}}, \ and\ \bibinfo {author}
  {\bibfnamefont {L.~H.}\ \bibnamefont {Tjeng}},\ }\href {\doibase
  10.1103/PhysRevB.96.155130} {\bibfield  {journal} {\bibinfo  {journal} {Phys.
  Rev. B}\ }\textbf {\bibinfo {volume} {96}},\ \bibinfo {pages} {155130}
  (\bibinfo {year} {2017})}\BibitemShut {NoStop}%
\bibitem [{Note1()}]{Note1}%
  \BibitemOpen
  \bibinfo {note} {Slater integrals ($F$ and $G$) and spin orbit interaction
  ($\zeta $) used for Sm$^{3+}$ 3$d^{10}$4$f^{5}$ and 3$d^{9}$4$f^{5}$ are
  ($F^{2}_{f}$, $F^{4}_{f}$, $F^{6}_{f}$, $\zeta _{f}$) = (13.64, 8.56, 6.16,
  0.144) and (15.31, 9.68, 6.98, 0.181), respectively, and ($F^{2}_{fd}$,
  $F^{4}_{fd}$, $G^{1}_{fd}$, $G^{3}_{fd}$,$G^{5}_{fd}$,$\zeta _{d}$) = (9.63,
  4.49, 6.87, 4.03, 2.78, 10.35)}\BibitemShut {NoStop}%
\bibitem [{\citenamefont {Suzuki}\ \emph {et~al.}(2000)\citenamefont {Suzuki},
  \citenamefont {Kawai}, \citenamefont {Takahashi}, \citenamefont {Vlaicu},
  \citenamefont {Adachi},\ and\ \citenamefont {Mukoyama}}]{Suzuki:2000aa}%
  \BibitemOpen
  \bibfield  {author} {\bibinfo {author} {\bibfnamefont {C.}~\bibnamefont
  {Suzuki}}, \bibinfo {author} {\bibfnamefont {J.}~\bibnamefont {Kawai}},
  \bibinfo {author} {\bibfnamefont {M.}~\bibnamefont {Takahashi}}, \bibinfo
  {author} {\bibfnamefont {A.-M.}\ \bibnamefont {Vlaicu}}, \bibinfo {author}
  {\bibfnamefont {H.}~\bibnamefont {Adachi}}, \ and\ \bibinfo {author}
  {\bibfnamefont {T.}~\bibnamefont {Mukoyama}},\ }\href {\doibase
  https://doi.org/10.1016/S0301-0104(99)00380-8} {\bibfield  {journal}
  {\bibinfo  {journal} {Chem. Phys.}\ }\textbf {\bibinfo {volume} {253}},\
  \bibinfo {pages} {27 } (\bibinfo {year} {2000})}\BibitemShut {NoStop}%
\bibitem [{\citenamefont {Mori}\ and\ \citenamefont
  {Tanemura}(2007)}]{Mori:2007aa}%
  \BibitemOpen
  \bibfield  {author} {\bibinfo {author} {\bibfnamefont {Y.}~\bibnamefont
  {Mori}}\ and\ \bibinfo {author} {\bibfnamefont {S.}~\bibnamefont
  {Tanemura}},\ }\href {\doibase https://doi.org/10.1016/j.apsusc.2006.08.011}
  {\bibfield  {journal} {\bibinfo  {journal} {Appl. Surf. Sci.}\ }\textbf
  {\bibinfo {volume} {253}},\ \bibinfo {pages} {3856 } (\bibinfo {year}
  {2007})}\BibitemShut {NoStop}%
\bibitem [{\citenamefont {Dallera}\ \emph {et~al.}(2005)\citenamefont
  {Dallera}, \citenamefont {Braicovich}, \citenamefont {Du{\`o}}, \citenamefont
  {Palenzona}, \citenamefont {Panaccione}, \citenamefont {Paolicelli},
  \citenamefont {Cowie},\ and\ \citenamefont {Zegenhagen}}]{Dallera:2005aa}%
  \BibitemOpen
  \bibfield  {author} {\bibinfo {author} {\bibfnamefont {C.}~\bibnamefont
  {Dallera}}, \bibinfo {author} {\bibfnamefont {L.}~\bibnamefont {Braicovich}},
  \bibinfo {author} {\bibfnamefont {L.}~\bibnamefont {Du{\`o}}}, \bibinfo
  {author} {\bibfnamefont {A.}~\bibnamefont {Palenzona}}, \bibinfo {author}
  {\bibfnamefont {G.}~\bibnamefont {Panaccione}}, \bibinfo {author}
  {\bibfnamefont {G.}~\bibnamefont {Paolicelli}}, \bibinfo {author}
  {\bibfnamefont {B.}~\bibnamefont {Cowie}}, \ and\ \bibinfo {author}
  {\bibfnamefont {J.}~\bibnamefont {Zegenhagen}},\ }\href {\doibase
  https://doi.org/10.1016/j.nima.2005.05.017} {\bibfield  {journal} {\bibinfo
  {journal} {Nucl. Instrum. Methods Phys. Res. A}\ }\textbf {\bibinfo {volume}
  {547}},\ \bibinfo {pages} {113 } (\bibinfo {year} {2005})}\BibitemShut
  {NoStop}%
\bibitem [{\citenamefont {Eguchi}\ \emph {et~al.}(2009)\citenamefont {Eguchi},
  \citenamefont {Okamoto}, \citenamefont {Hiroi}, \citenamefont {Shin},
  \citenamefont {Chainani}, \citenamefont {Tanaka}, \citenamefont {Matsunami},
  \citenamefont {Takata}, \citenamefont {Nishino}, \citenamefont {Tamasaku},
  \citenamefont {Yabashi},\ and\ \citenamefont {Ishikawa}}]{Eguchi:2009aa}%
  \BibitemOpen
  \bibfield  {author} {\bibinfo {author} {\bibfnamefont {R.}~\bibnamefont
  {Eguchi}}, \bibinfo {author} {\bibfnamefont {Y.}~\bibnamefont {Okamoto}},
  \bibinfo {author} {\bibfnamefont {Z.}~\bibnamefont {Hiroi}}, \bibinfo
  {author} {\bibfnamefont {S.}~\bibnamefont {Shin}}, \bibinfo {author}
  {\bibfnamefont {A.}~\bibnamefont {Chainani}}, \bibinfo {author}
  {\bibfnamefont {Y.}~\bibnamefont {Tanaka}}, \bibinfo {author} {\bibfnamefont
  {M.}~\bibnamefont {Matsunami}}, \bibinfo {author} {\bibfnamefont
  {Y.}~\bibnamefont {Takata}}, \bibinfo {author} {\bibfnamefont
  {Y.}~\bibnamefont {Nishino}}, \bibinfo {author} {\bibfnamefont
  {K.}~\bibnamefont {Tamasaku}}, \bibinfo {author} {\bibfnamefont
  {M.}~\bibnamefont {Yabashi}}, \ and\ \bibinfo {author} {\bibfnamefont
  {T.}~\bibnamefont {Ishikawa}},\ }\href {\doibase 10.1063/1.3086666}
  {\bibfield  {journal} {\bibinfo  {journal} {J. Appl. Phys.}\ }\textbf
  {\bibinfo {volume} {105}},\ \bibinfo {pages} {056103} (\bibinfo {year}
  {2009})}\BibitemShut {NoStop}%
\bibitem [{\citenamefont {Yamamoto}\ \emph {et~al.}(2018)\citenamefont
  {Yamamoto}, \citenamefont {Ootsuki}, \citenamefont {Shimonaka}, \citenamefont
  {Shibata}, \citenamefont {Kodera}, \citenamefont {Okawa}, \citenamefont
  {Saitoh}, \citenamefont {Horio}, \citenamefont {Fujimori}, \citenamefont
  {Kumigashira}, \citenamefont {Ono}, \citenamefont {Ikenaga}, \citenamefont
  {Miyasaka}, \citenamefont {Tajima},\ and\ \citenamefont
  {Yoshida}}]{Yamamoto:2018aa}%
  \BibitemOpen
  \bibfield  {author} {\bibinfo {author} {\bibfnamefont {S.}~\bibnamefont
  {Yamamoto}}, \bibinfo {author} {\bibfnamefont {D.}~\bibnamefont {Ootsuki}},
  \bibinfo {author} {\bibfnamefont {D.}~\bibnamefont {Shimonaka}}, \bibinfo
  {author} {\bibfnamefont {D.}~\bibnamefont {Shibata}}, \bibinfo {author}
  {\bibfnamefont {K.}~\bibnamefont {Kodera}}, \bibinfo {author} {\bibfnamefont
  {M.}~\bibnamefont {Okawa}}, \bibinfo {author} {\bibfnamefont
  {T.}~\bibnamefont {Saitoh}}, \bibinfo {author} {\bibfnamefont
  {M.}~\bibnamefont {Horio}}, \bibinfo {author} {\bibfnamefont
  {A.}~\bibnamefont {Fujimori}}, \bibinfo {author} {\bibfnamefont
  {H.}~\bibnamefont {Kumigashira}}, \bibinfo {author} {\bibfnamefont
  {K.}~\bibnamefont {Ono}}, \bibinfo {author} {\bibfnamefont {E.}~\bibnamefont
  {Ikenaga}}, \bibinfo {author} {\bibfnamefont {S.}~\bibnamefont {Miyasaka}},
  \bibinfo {author} {\bibfnamefont {S.}~\bibnamefont {Tajima}}, \ and\ \bibinfo
  {author} {\bibfnamefont {T.}~\bibnamefont {Yoshida}},\ }\href {\doibase
  10.7566/JPSJ.87.024708} {\bibfield  {journal} {\bibinfo  {journal} {J. Phys.
  Soc. Jpn}\ }\textbf {\bibinfo {volume} {87}},\ \bibinfo {pages} {024708}
  (\bibinfo {year} {2018})}\BibitemShut {NoStop}%
\bibitem [{\citenamefont {Horio}\ \emph {et~al.}(2018)\citenamefont {Horio},
  \citenamefont {Krockenberger}, \citenamefont {Yamamoto}, \citenamefont
  {Yokoyama}, \citenamefont {Takubo}, \citenamefont {Hirata}, \citenamefont
  {Sakamoto}, \citenamefont {Koshiishi}, \citenamefont {Yasui}, \citenamefont
  {Ikenaga}, \citenamefont {Shin}, \citenamefont {Yamamoto}, \citenamefont
  {Wadati},\ and\ \citenamefont {Fujimori}}]{Horio:2018aa}%
  \BibitemOpen
  \bibfield  {author} {\bibinfo {author} {\bibfnamefont {M.}~\bibnamefont
  {Horio}}, \bibinfo {author} {\bibfnamefont {Y.}~\bibnamefont
  {Krockenberger}}, \bibinfo {author} {\bibfnamefont {K.}~\bibnamefont
  {Yamamoto}}, \bibinfo {author} {\bibfnamefont {Y.}~\bibnamefont {Yokoyama}},
  \bibinfo {author} {\bibfnamefont {K.}~\bibnamefont {Takubo}}, \bibinfo
  {author} {\bibfnamefont {Y.}~\bibnamefont {Hirata}}, \bibinfo {author}
  {\bibfnamefont {S.}~\bibnamefont {Sakamoto}}, \bibinfo {author}
  {\bibfnamefont {K.}~\bibnamefont {Koshiishi}}, \bibinfo {author}
  {\bibfnamefont {A.}~\bibnamefont {Yasui}}, \bibinfo {author} {\bibfnamefont
  {E.}~\bibnamefont {Ikenaga}}, \bibinfo {author} {\bibfnamefont
  {S.}~\bibnamefont {Shin}}, \bibinfo {author} {\bibfnamefont {H.}~\bibnamefont
  {Yamamoto}}, \bibinfo {author} {\bibfnamefont {H.}~\bibnamefont {Wadati}}, \
  and\ \bibinfo {author} {\bibfnamefont {A.}~\bibnamefont {Fujimori}},\ }\href
  {\doibase 10.1103/PhysRevLett.120.257001} {\bibfield  {journal} {\bibinfo
  {journal} {Phys. Rev. Lett.}\ }\textbf {\bibinfo {volume} {120}},\ \bibinfo
  {pages} {257001} (\bibinfo {year} {2018})}\BibitemShut {NoStop}%
\bibitem [{\citenamefont {Lee}\ \emph {et~al.}(2014)\citenamefont {Lee},
  \citenamefont {Vaezi}, \citenamefont {Fischer},\ and\ \citenamefont
  {Kim}}]{Lee:2014aa}%
  \BibitemOpen
  \bibfield  {author} {\bibinfo {author} {\bibfnamefont {K.}~\bibnamefont
  {Lee}}, \bibinfo {author} {\bibfnamefont {A.}~\bibnamefont {Vaezi}}, \bibinfo
  {author} {\bibfnamefont {M.~H.}\ \bibnamefont {Fischer}}, \ and\ \bibinfo
  {author} {\bibfnamefont {E.-A.}\ \bibnamefont {Kim}},\ }\href {\doibase
  10.1103/PhysRevB.90.214510} {\bibfield  {journal} {\bibinfo  {journal} {Phys.
  Rev. B}\ }\textbf {\bibinfo {volume} {90}},\ \bibinfo {pages} {214510}
  (\bibinfo {year} {2014})}\BibitemShut {NoStop}%
\bibitem [{\citenamefont {Fu}\ and\ \citenamefont {Kane}(2008)}]{Fu:2008aa}%
  \BibitemOpen
  \bibfield  {author} {\bibinfo {author} {\bibfnamefont {L.}~\bibnamefont
  {Fu}}\ and\ \bibinfo {author} {\bibfnamefont {C.~L.}\ \bibnamefont {Kane}},\
  }\href {\doibase 10.1103/PhysRevLett.100.096407} {\bibfield  {journal}
  {\bibinfo  {journal} {Phys. Rev. Lett.}\ }\textbf {\bibinfo {volume} {100}},\
  \bibinfo {pages} {096407} (\bibinfo {year} {2008})}\BibitemShut {NoStop}%
\end{thebibliography}%

\end{document}